\documentclass[twocolumn]{el-author}

\usepackage{algorithm} 
\usepackage{algorithmic} 

\newcommand{\ua}{\uparrow}
\newcommand{\nc}{\newcommand}
\nc{\da}{\downarrow} \nc{\hc}{\hat{c}} \nc{\hS}{\hat{S}}
\nc{\bra}{\langle} \nc{\ket}{\rangle} \nc{\eq}{equation (\ref}
\nc{\h}{\hat} \nc{\hT}{\h{T}}\nc{\be}{\begin{eqnarray}}
\nc{\ee}{\end{eqnarray}}\nc{\rd}{\textrm{d}}\nc{\e}{eqnarray}\nc{\hR}{\hat{R}}\nc{\Tr}{\mathrm{Tr}}
\nc{\tS}{\tilde{S}}\nc{\tr}{\mathrm{tr}}\nc{\8}{\infty}\nc{\lgs}{\bra\ua,\phi|}\nc{\rgs}{|\ua,\phi\ket}
\nc{\hU}{\hat{U}}\nc{\lfs}{\bra\phi|}\nc{\rfs}{|\phi\ket}\nc{\hZ}{\hat{Z}}\nc{\hd}{\hat{d}}\nc{\mD}{\mathcal{D}}
\nc{\bd}{\bar{d}}\nc{\bc}{\bar{c}}\nc{\mc}{\mathcal}\nc{\ea}{eqnarray}\nc{\mG}{\mathcal{G}}\nc{\bce}{\begin{center}}
\nc{\ece}{\end{center}}
\date{3th October 2014}
\begin{document}

\title{Spatially correlated channel estimation based on block iterative support detection for large-scale MIMO}

\author{Wenqian Shen, Linglong Dai, Zhen Gao, and Zhaocheng Wang, \textit{IET Fellow}}

\abstract{Downlink channel estimation with low pilot overhead is an
important and challenging problem in large-scale MIMO systems due to
the substantially increased MIMO channel dimension. In this letter,
we propose a block iterative support detection (block-ISD) based
algorithm for downlink channel estimation to reduce the pilot
overhead, which is achieved by fully exploiting the block sparsity
inherent in the block-sparse equivalent channel derived from the
spatial correlations of MIMO channels. Furthermore, unlike
conventional compressive sensing (CS) algorithms that rely on prior
knowledge of the sparsity level, block-ISD relaxes this demanding
requirement and is thus more practically appealing. Simulation
results demonstrate that block-ISD yields better normalized mean
square error (NMSE) performance than classical CS algorithms, and
achieve a reduction of $84\%$ pilot overhead than conventional
channel estimation techniques.}

\maketitle

\section{Introduction}
Recently, large-scale MIMO with a large number of antennas at the
base station (BS) has emerged as a key promising technology for
future 5G wireless communications. It is proved that large-scale
MIMO can reduce transmit power as well as increase spectrum
efficiency by orders of magnitude \cite{scaling_up_mimo}. In such
systems, accurate downlink channel state information (CSI) is
essential for channel adaptive techniques such as water-filling,
beamforming, etc. As the number of antennas keeps increasing in
large-scale MIMO systems, efficient low-overhead channel estimation
will be an increasingly important and challenging problem.
Conventional channel estimation techniques including least square
(LS)  and minimum mean square error (MMSE) \cite{Ozdemir} are not
suitable for large-scale MIMO systems due to the number of required
orthogonal pilots scales linearly with the number of antennas at the
BS, which results in prohibitively high pilot overhead. Recently,
several efficient channel estimation schemes based on compressive
sensing (CS) have been proposed to reduce pilot overhead by taking
channel sparsity (i.e., the channel is sparse) into account [3-5].
However, these CS-based channel estimation schemes usually assume
prior knowledge of the sparsity level, i.e., the number of non-zero
elements of the channel impulse response (CIR), which is usually
unknown and difficult to be accurately estimated in practice. In
addition, due to the physical propagation characteristics of
multiple antennas, CIRs associated with different antennas
inevitably share spatial correlations \cite{CommonSupport}, which
have not been considered by the existing CS-based channel estimation
schemes to further improve the performance.

Building upon the iterative support detection (ISD) algorithm
\cite{ISD}, in this letter we propose an improved block iterative
support detection (block-ISD) based algorithm for downlink channel
estimation to reduce the pilot overhead. Specifically, by taking
into account the spatial correlations of MIMO channels caused by the
physical propagation characteristics of multiple antennas and close
antenna spacing at the BS, we generate the block-sparse equivalent
CIR with the preferred block sparsity. Accordingly, we propose a
block-ISD based algorithm to exploit the block sparsity to
significantly reduce the pilot overhead required for accurate
channel estimation. Note that block-ISD requires no prior knowledge
of the sparsity level, and is thus more practically appealing than
conventional CS-based algorithms.

\section{System model}
We consider a large-scale MIMO system with $N_T$ antennas at the BS
and $K$ scheduled single-antenna users ($N_T>>K$) using the commonly
used OFDM modulation. Normally, frequency-domain pilots are used for
channel estimation in OFDM-based systems [3-5]. At the user side,
the received pilots in the frequency domain can be expressed as
\begin{equation}
\mathbf{y}_\Omega=\sum_{i=1}^{N_T}\mathbf{C}_i(\mathbf{F}_L)_\Omega\mathbf{h}_i+\mathbf{n}_\Omega,
\end{equation}
where $\Omega$ is the index set of subcarriers assigned to pilots,
which can be randomly selected from the subcarrier set
$[1,2,\cdots,N]$. $\mathbf{C}_i=\text{diag}\{\mathbf{c}_i\}$ with
$\mathbf{c}_i\in\mathcal{C}^{p\times 1}$ being the pilot vector for
the $i$th transmit antenna, and the number of pilots is $p$.
$\mathbf{F}_L\in\mathcal{C}^{N\times L}$ is a sub-matrix consisting
of the first $L$ columns of the discrete fourier transform (DFT)
matrix of size $N\times N$, where $N$ is the OFDM symbol length.
$(\mathbf{F}_L)_\Omega$ is the sub-matrix consisting of the rows of
$\mathbf{F}_L$ with indexes from $\Omega$.
$\mathbf{h}_i=[\mathbf{h}_i(1), \mathbf{h}_i(2),\cdots,
\mathbf{h}_i(L)]^T$ denotes the CIR between the $i$th transmit
antenna of the BS and the single receive antenna of the user with
the maximum channel length $L$. $\mathbf{n}_\Omega=[n_1, \cdots,
n_p]^T$ represents the noise vector consisting of independent and
identically distributed (i.i.d.) additive white complex Gaussian
noise (AWGN) variables with zero mean and unit-variance. For
simplicity, (1) can also be written as
\begin{equation}
\mathbf{y}_\Omega=\mathbf{P}\mathbf{h}+\mathbf{n}_\Omega,
\end{equation}
where $\mathbf{P}=[\mathbf{C}_1(\mathbf{F}_L)_\Omega,\mathbf{C}_2(\mathbf{F}_L)_\Omega,\cdots, \mathbf{C}_{N_T}(\mathbf{F}_L)_\Omega]$ can be regarded as the sensing matrix, $\mathbf{h}=[\mathbf{h}_1^T,\mathbf{h}_2^T,\cdots, \mathbf{h}_{N_T}^T]^T$ is the aggregate CIR from $N_T$ antennas to be estimated in large-scale MIMO systems.

\section{Downlink channel estimation based on block-ISD}
In this section, we firstly generate the block-sparse equivalent CIR
by considering spatial correlations of CIRs associated with
different antennas. Then, to fully exploit the block sparsity
inherent in the block-sparse equivalent CIR, we propose an improved
block-ISD algorithm for channel estimation.

\section{1. Generation of the block-sparse equivalent CIR}
Due to the physical propagation characteristics of multiple antennas
and close antenna spacing at the BS, CIRs
$\{\mathbf{h}_i\}_{i=1}^{N_T}$ associated with different transmit
antennas have similar path arrival times, and thus they share a
common support \cite{CommonSupport}, i.e.,
$\mathbf{\Gamma}_{\mathbf{h}_1}=\mathbf{\Gamma}_{\mathbf{h}_2}=\cdots
=\mathbf{\Gamma}_{\mathbf{h}_{N_T}}$, where
$\mathbf{\Gamma}_{\mathbf{h}_i}=\{k:\mathbf{h}_i(k)\neq 0\}$ denotes
the support of $\mathbf{h}_i$. Since the CIRs from different
transmit antennas share a common support, we can group the elements
of $\mathbf{h}_i$ with the same indexes into non-zero blocks and
zero blocks to generate the block-sparse equivalent CIR
$\mathbf{g}=[\mathbf{g}_1, \mathbf{g}_2, \cdots, \mathbf{g}_L]^T$.
More specifically, the relationship between $\mathbf{h}$ and
$\mathbf{g}$ can be denoted by
\begin{equation}
\mathbf{g}((l-1)N_T+n_t)=\mathbf{h}((n_t-1)L+l),
\end{equation}
where $l=1, 2, \cdots, L$ and $n_t=1, 2, \cdots, N_T$. It is
important that if we equally divide $\mathbf{g}$ into $L$ blocks
with $N_T$ elements in each block, these $N_T$ continuous elements
in the $l$th block $\mathbf{g}_l$ are all zeros or non-zeros. Thus,
the generated block-sparse equivalent CIR $\mathbf{g}$ enjoys the
preferred property of {\it  block sparsity}. This implies that we
can treat the $N_T$ continuous elements of the support
$\mathbf{\Gamma}_\mathbf{g}$ of $\mathbf{g}$ as a whole and update
them simultaneously.

Accordingly, similar to (3), we can obtain a new sensing matrix
$\mathbf{\Theta}$ by rearranging the columns of $\mathbf{P}$ in (2)
as
 \begin{equation}
\mathbf{\Theta}(:,(l-1)N_T+n_t)=\mathbf{P}(:,(n_t-1)L+l).
\end{equation}
Therefore, the channel estimation problem (2) can be reformulated as
\begin{equation}
\mathbf{y}_\Omega=\mathbf{\Theta}\mathbf{g}+\mathbf{n}_\Omega.
\end{equation}
This is an underdetermined problem with $\mathbf{g}$ of size
$N_{T}L\times 1$ and $\mathbf{y}_\Omega$ of size $p\times 1$, where
$p$ is usually much smaller than $N_{T}L$ due to the large number of
antennas and the limited pilot overhead. Traditional channel
estimation techniques such as LS and MMSE can not recover
$\mathbf{g}$ with limited pilot overhead. In this letter, the block
sparsity inherent in the generated block-sparse equivalent CIR
$\mathbf{g}$ will be utilized by the proposed block-ISD algorithm in
the following section to solve this problem.

\section{2. Downlink channel estimation based on block-ISD}
\vspace{+4mm}
\begin{algorithm}[tb]
\renewcommand{\algorithmicrequire}{\textbf{Input:}}
\renewcommand\algorithmicensure {\textbf{Output:} }
\caption{Block-ISD Algorithm}
\label{Block-ISD algorithm}
\begin{algorithmic}[1]
\REQUIRE ~~\\
1) Measurements $\mathbf{y}_{\Omega}$ \\
2) Sensing matrix $\mathbf{\Theta}$\\
\STATE Initialization: \\
$s=0$ and $\mathbf{\Gamma}_{\mathbf{g}}^{(0)}=\emptyset$ \WHILE
{$\text{Card}(\mathbf{\Gamma}_{\mathbf{g}}^{(s)})<N_TL-p$} \STATE
$W^{(s)}=(\mathbf{\Gamma_{\mathbf{g}}^{(s)}})^c$; \STATE
$\mathbf{g}^{(s)}\leftarrow\min_{\mathbf{g}^{(s)}}\|\mathbf{g}^{(s)}_{W^{(s)}}\|_1\;
\hspace{+3mm} s.t. \;
\mathbf{y}_\Omega=\mathbf{\Theta}\mathbf{g}^{(s)}+\mathbf{n}_\Omega$;
\STATE $\mathbf{v}^{(s)}=\text{Sort}(\mathbf{g}^{(s)})$; \STATE
$i\leftarrow \min _i\; \hspace{+1.5mm} s.t.
|\mathbf{v}^{(s)}(i+1)|-|\mathbf{v}^{(s)}(i)|>|\tau^{(s)}|$; \STATE
$\epsilon^{(s)}=|\mathbf{v}^{(s)}(i)|$; \STATE
$\mathbf{\Gamma}_{\mathbf{v}}^{(s)}=\{k\;\hspace{+1.5mm} s.t.
\;|\mathbf{v}^{(s)}(k)|>\epsilon^{(s)}\}$; \STATE
$\mathbf{\Gamma}_{\mathbf{g}}^{(s)}=\{(l-1)N_T+1:1:lN_T\;\hspace{+1.5mm}s.t.
\;\text{Card}(\{(l-1)N_T+1:1:lN_T\}\cap\mathbf{\Gamma_{\mathbf{v}}^{(s)}})>N_T/2\}$;
\STATE $s=s+1$. \ENDWHILE \RETURN
$\mathbf{\hat{g}}=\mathbf{g}^{(s)}$
\ENSURE ~~\\
Recovered block-sparse equivalent CIR $\mathbf{\hat{g}}$\\
\end{algorithmic}
\end{algorithm}
The pseudocode of block-ISD is in \textbf{Algorithm 1}. Note that
block-ISD updates all the elements of the recovered signal
$\mathbf{g}^{(s)}$ in the $s$th iteration through solving the
truncated basic pursuit (BP) problem \cite{BP} in step 4:
\begin{equation}
\min_{\mathbf{g}^{(s)}}\|\mathbf{g}^{(s)}_{W^{(s)}}\|_1\;
\hspace{+3mm} s.t. \;
\mathbf{y}_\Omega=\mathbf{\Theta}\mathbf{g}^{(s)}+\mathbf{n}_\Omega,
\end{equation}
where $\|\mathbf{g}^{(s)}_{W^{(s)}}\|_1=\sum_{w\in
W^{(s)}}|\mathbf{g}^{(s)}(w)|$. This problem can be efficiently
solved by calling a BP algorithm such as YALL1 \cite{ISD}. Then, the
support $\mathbf{\Gamma}_{\mathbf{g}}^{(s)}$ is updated in the $s$th
iteration through the adjacent support detection in steps 5-9. In
these five steps, we firstly sort $\mathbf{g}^{(s)}$ in an ascending
order in step 5 to obtain $\mathbf{v}^{(s)}$. Then, the support of
$\mathbf{v}^{(s)}$ can be detected based on the `first significant
jump' rule \cite{ISD} in step 6, which searches for the smallest $i$
that satisfies
$|\mathbf{v}^{(s)}(i+1)|-|\mathbf{v}^{(s)}(i)|>|\tau^{{s}}|$, where
$\tau^{(s)}=(LN_T)^{-1}\|\mathbf{v}^{(s)}\|_\infty$ \cite{ISD}. The
smallest $i$ is the index where the `first significant jump' occurs
in an ascending ordered vector $\mathbf{v}^{(s)}$. Next we set the
threshold $\epsilon^{(s)}=|\mathbf{v}^{(s)}(i)|$ in step 7, then the
support of $\mathbf{v}^{(s)}$ can be updated in step 8 based on this
threshold. Finally, due to the block sparsity of $\mathbf{g}^{(s)}$,
the support of $\mathbf{g}^{(s)}$ can be updated in step 9, where
$\text{Card}(\cdot)$ denotes the number of elements of a set.

Note that the support $\mathbf{\Gamma}_{\mathbf{g}}^{(s)}$ is
independent of $\mathbf{\Gamma}_{\mathbf{g}}^{(s-1)}$ in block-ISD,
which is different from the classical greedy  CS algorithm called
orthogonal matching pursuit (OMP) \cite{Stru_CS_Theory_Appli}. In
OMP, only one element of $\mathbf{\Gamma}_{\mathbf{g}}^{(s)}$ is
updated in each iteration, and once an element is added to
$\mathbf{\Gamma}_{\mathbf{g}}^{(s)}$, this element will not be
removed in the following iterations. From this aspect, block-ISD is
similar to subspace pursuit (SP) \cite{SubspacePursuit} and
compressive sampling matching pursuit (CoSaMP)
\cite{Stru_CS_Theory_Appli}. They update all elements of the
recovered signal in every iteration, whereby the support detection
not only selects the desired elements but also removes the undesired
elements. However, the support detection of SP and CoSaMP is based
on the sparsity level assumed to be known as a priori, while the
support detection of block-ISD is based on the sparsity-independent
threshold $\epsilon^{(s)}$. Thus, block-ISD can recover the signal
without prior knowledge of the channel sparsity level.

The key difference between ISD and block-ISD is the consideration of
the block sparsity of $\mathbf{g}^{(s)}$. For a certain non-zero
block of $\mathbf{g}$, the continuous $N_T$ elements of this block
are supposed to be all non-zeros, so their indexes are supposed to
be included in $\mathbf{\Gamma}_{\mathbf{g}}^{(s)}$. However, some
indexes may be incorrectly detected due to the impact of noise.
Nevertheless, we can determine whether this block is a zero block or
a non-zero block by comparing the number of indexes included in
$\mathbf{\Gamma}_{\mathbf{g}}^{(s)}$ with $N_T/2$ (half of the block
length $N_T$) in step 9. Only when more than half of the indexes of
a certain block are included in
$\mathbf{\Gamma}_{\mathbf{g}}^{(s)}$, then all $N_T$ indexes of this
block will be added in $\mathbf{\Gamma}_{\mathbf{g}}^{(s)}$. This
mechanism considering the block sparsity is expected to increase the
robustness of the support detection and thus improve the channel
estimation performance as will be verified by the simulation results
in the following section.

\section{Simulation results}
Simulations have been conducted to validate the performance of
block-ISD. We consider an $N_T=32$ large-scale MIMO system with the
system bandwidth of 50MHz and the OFDM symbol length $N=4096$. We
adopt the ITU Vehicular-A channel model \cite{Dai13} with the
maximum channel length $L=128$. Fig. 1 shows the normalized mean
square error (NMSE) performance comparison between the proposed
block-ISD and the classical ISD [7] and BP [8] algorithms, where the
number of pilots is $p=640$. In addition, the performance of the
exact least square (LS) algorithm assuming the exact knowledge of
the support of block-sparse equivalent CIR is also presented as the
lower bound of NMSE for comparison. It can be observed that
block-ISD outperforms both classical ISD and BP algorithms.
Specifically, block-ISD achieves more than 4 dB SNR gain than ISD
and BP algorithms when the target NMSE of $10^{-1}$ is considered.
The performance gain is mainly attributed to the exploration of
block sparsity inherent in the generated block-sparse equivalent
CIR. Note that block-ISD obviously outperforms ISD when SNR is not
very high. This is due to the fact that block-ISD is more capable of
correcting the support detection error caused by the additive noise
than ISD when SNR is not very high (e.g., SNR < 20 dB), which
therefore enhances the support detection and ultimately leads to a
lower NMSE. For conventional channel estimation techniques such as
LS and MMSE \cite{Ozdemir}, the number of pilot $p$ should be as
large as $LN_T=128\times 32=4096$ to ensures (5) as an
overdetermined problem. That is to say, block-ISD achieves a
substantial reduction of $(4096-640)/4096=84\%$ pilot overhead
compared with these conventional channel estimation techniques
without considering the channel sparsity.

\begin{figure} \includegraphics[width=88mm]{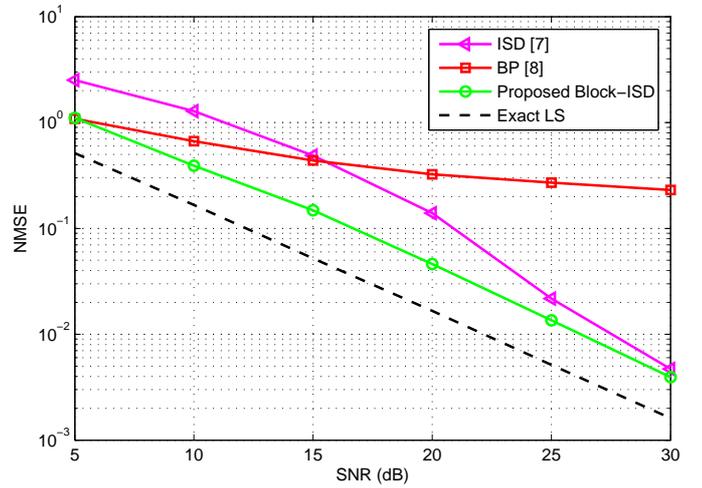}
\caption{NMSE performance comparison between block-ISD, ISD, and
BP.}
\end{figure}

\section{Conclusion}
In this letter, we have investigated the challenging problem of
downlink channel estimation with acceptable pilot overhead for
large-scale MIMO systems. It is found that by exploring the block
sparsity inherent in the block-sparse equivalent CIR, which is
generated by considering the spatial correlations of MIMO channels,
the proposed block-ISD algorithm could improve the channel
estimation performance by more than 4 dB than classical ISD and BP
algorithms. In addition, we have shown that block-ISD requires no
prior knowledge of the channel sparsity level, thereby making an
important step toward practical implementation. Simulation results
have demonstrated that block-ISD can achieve a reduction of $84\%$
pilot overhead than conventional channel estimation techniques. The
extension of block sparsity to temporally correlated channels will
be left as future work.

 \vskip3pt \ack{This work was supported by National Key Basic
Research Program of China (Grant No. 2013CB329203)}

\vskip5pt

\noindent Wenqian Shen, Linglong Dai, Zhen Gao, and Zhaocheng Wang (\textit{Tsinghua National Laboratory for Information Science and Technology, Department of Electronic Engineering, Tsinghua University, Beijing 100084, China})
\vskip3pt

\noindent E-mail: daill@tsinghua.edu.cn

\end{document}